\documentstyle[12pt,preprint,aps]{revtex}

\begin{document}
\title
{Andreev reflection in Si-engineered Al/InGaAs hybrid junctions 
}
\author{Silvano De Franceschi, Francesco Giazotto, 
and Fabio Beltram}
\address{Scuola Normale Superiore and INFM, I-56126 Pisa,
Italy}
\author{Lucia Sorba$^{a)}$, Marco Lazzarino, and Alfonso Franciosi$^{b)}$}
\address{Laboratorio Nazionale TASC-INFM, Area di Ricerca, Padriciano
99, I-34012 Trieste, Italy}
\footnotetext[1]{Also with Istituto ICMAT-CNR, Via Salaria km 29.3,
I-00016 Monterotondo, Italy.}
\footnotetext[2]{Also with Universit\`a di Trieste, Trieste, Italy.}
\maketitle
\begin{abstract}

Andreev-reflection dominated transport is demonstrated in 
Al/n-In$_{0.38}$Ga$_{0.62}$As superconductor-semiconductor 
junctions grown by molecular-beam epitaxy on GaAs(001). 
High junction transparency was achieved in low-doped devices by 
exploiting Si interface bilayers to suppress the native Schottky barrier.
It is argued that this technique is ideally suited for the fabrication
of ballistic transport hybrid microstructures.

\end{abstract}

Superconductor-semiconductor (S-Sm) hybrid junctions are
attracting increasing attention as their
potential for the realization of exotic electronic-state
and transport property configurations becomes more and more
apparent \cite{been,klei,lamb}. In particular, merging ideas and techniques
developed for semiconductor nanostructures
with those of superconductivity research is proving extremely
fruitful although several challenges still remain  
from the fabrication and theoretical standpoints.

Junction transparency
is the key to access the novel transport regimes of interest. 
In particular, it is a crucial  requirement for the observation of  
Andreev-reflection \cite{andreev}.
(In this process an electron injected from the Sm side condenses
into a Cooper pair in the S part of the junction with the simultaneous
retroreflection of a  hole along the electron time-reversed path.)
To achieve this high transparency different techniques have been
explored including metal deposition immediately after 
As-decapping \cite {kast}, 
Ar$^+$ back-sputtering \cite {nguy}, and {\it in situ} metallization in 
the molecular-beam epitaxy (MBE) chamber \cite{akaz}.
All these tests were performed in InAs-based S-Sm junctions where 
interface contamination is the main transparency-limiting factor.  
For the case of more established semiconductor materials (such as those
grown on either GaAs or InP) the strongest limitation arises
from the existence of a native Schottky barrier.  Here, 
penetrating contacts \cite{gao,williams} and heavily doped surface layers 
\cite{kast,tabo} were used to enhance junction transparency.  

In a recent study we have reported on a new technique \cite{silv}, 
alternative to doping, to obtain Schottky-barrier-free 
Al/n-In$_x$Ga$_{1-x}$As(001) junctions ($x \agt 0.3$) by MBE growth. 
This is based on the inclusion of an ultrathin Si interface layer 
under As flux which 
changes the pinning position of the Fermi level at the metal-semiconductor 
junction. This leads  to the total suppression of the Schottky barrier. 
In this Letter we demonstrate that this technique 
can be successfully employed to achieve high transparency in MBE-grown 
S-Sm hybrid junctions 
involving low-doped and low-In-content InGaAs alloys that are ideal 
candidates for the implementation of ballistic-transport structures. 

Al/n-In$_{0.38}$Ga$_{0.62}$As S-Sm junctions incorporating Si 
interface layers were 
grown by MBE at the TASC-INFM facility. Their schematic structure 
is shown in Fig. 1.  The semiconductor portion consists of a 
300-nm-thick GaAs buffer layer grown at 600 $^\circ$C on n-type GaAs(001) 
and Si-doped  at $n \sim 10^{18}$ cm$^{-3}$ followed by 
a 2-$\mu$m-thick n-In$_{0.38}$Ga$_{0.62}$As layer grown at
500 $^\circ$C with an inhomogeneous doping profile. 
The top 1.5-$\mu$m-thick region was doped at $n=6.5 \cdot 10^{16}$ 
cm$^{-3}$, 
the bottom buffer region (0.5 $\mu$m thick) was
heavily doped at $n \sim 10^{18}$ cm$^{-3}$.
After In$_{0.38}$Ga$_{0.62}$As growth the substrate temperature
was lowered to 300$^\circ$C and a Si atomic bilayer was deposited
under As flux \cite{silv}.
Al deposition was carried out {\it in situ} at room temperature.

Rectangular 100$\times$160 $\mu$m$^2$ Al/n-In$_{0.38}$Ga$_{0.62}$As 
junctions were patterned on the sample surface using standard 
photolithographic techniques and wet chemical etching. 
Two additional 100$\times$50 $\mu$m$^2$-wide and 200-nm-thick Au pads
were electron-beam evaporated just on top of every Al pattern in order to allow
four-wire electrical measurements.  
The sample was mounted on a non magnetic dual-in-line sample
holder, and 25-$\mu$m-thick gold
wires were connected to the gold pads by standard ultrasonic bonding technique.
Current-voltage (I--V) characterizations as a function of temperature
($T$) and static magnetic field ($H$)   
were performed in a $^3$He closed-cycle cryostat. 

The critical temperature ($T_c$) of the Al film was measured to be 
1.1 K, which corresponds to a superconducting gap $\Delta \approx 0.16$ meV.  
The normal-state resistance $R_N$ of our devices was 0.2 $\Omega$,
including the series-resistance contribution ($\approx 0.1 \Omega$) 
of the semiconductor. At $H=0$ and  below $T_c$, 
dc I--V characteristics exhibited important deviations from linearity
around zero bias.
These features can be better analyzed by plotting  the 
differential conductance ($G$) as a function of the applied bias ($V$).
In Fig. 2 we show a tipical set of $G$-$vs$-$V$ curves obtained at
different temperatures in the 0.33--1.03 K range. 
These data clearly show that the transport properties of our system 
are quite unlike those of a S-Sm tunnel junction. In fact even at $T=0.33$ K,
i.e. well below $T_c$, a high value of $G$ is observed at zero bias. 
At such low temperatures and at low bias (i.e., when the voltage drop
across the junction is lower than $\Delta /e$ \cite{nota}), 
transport is dominated by Andreev reflection. 
The observation of such pronounced Andreev reflection demonstrates
high junction transparency. The latter can be quantified in terms of a  
dimensionless parameter $Z$ 
according to the Blonder-Tinkham-Klapwijk (BTK) model \cite{btk} 
(in this model $Z$ is related to the normal-state transmission coefficient  
$t$ by $t=(1+Z^2)^{-1}$).  
This approach was developed in the contex of ballistic systems, but
has been widely applied to diffusive systems (like the present one)
in order to gain an estimate of the junction trasmissivity 
\cite{gao,tabo,van,klei}. 
To analyze the data of Fig. 2 we followed the model 
by Chaudhuri and Bagwell \cite{chau},  which is the three-dimensional
generalization of the BTK model.
For our S-Sm junction we found $Z \approx 1$ corresponding
to a $\sim$50 \% normal-state transmission coefficient.  
We note that without the aid of the 
Si-interface-layer technique, doping concentrations 
over two orders of magnitude greater then that employed here 
would be necessary to achieve comparable transmissivity 
(see e.g. Refs. \cite{kast,gao,tabo}).
This drastic reduction in the needed impurity concentration is a very
attractive feature for the fabrication of ballistic structures.     
It should also be noted that our reported $Z$
value is close to the intrinsic transmissivity limit 
related to the Fermi-velocity mismatch between Al and InGaAs
\cite{BT}. 

We should also like to comment on the homogeneity of our junctions.
Our estimate of $Z$ leads to a  
theoretical value of the normal-state resistance $R_N^{th}$   
which is much smaller than the experimental value $R_N^{exp}$:
$R_N^{th}/R_N^{exp}=0.003$ . This suggests that 
only a small fraction  ($R_N^{th}/R_N^{exp}$) of the contact area has the 
high transparency and dominates the transport properties of the junction,
as already reported by other authors with different fabrication
techniques \cite{gao,van}. Values of $R_N^{th}/R_N^{exp}$ ranging from
$\sim 10^{-4}$ to $\sim 10^{-2}$ can be found in the literature 
(see, e.g., Refs. \cite{kast,gao,van}).

Such inhomogeneities, however, are not perceptible on the lateral 
length scale of our contacts and we observed a high uniformity in the transport 
properties of all junctions studied. 

The superconducting nature of the conductance dip for $|V|<\Delta/e$
is proved by its pronounced dependence on temperature and magnetic
field. Figure 2 shows how the zero-bias differential-conductance dip 
observed at $T=0.33$ K progressively weakens for $T$ approaching $T_c$.
This fact is consistent with the well-known
temperature-induced suppression of the superconducting energy gap 
$\Delta$. Far from $V=0$
the conductance is only marginally affected by temperature as
expected for a S-Sm junction when $|V|$ is significantly
larger than $\Delta/e$ \cite{btk}. 
A small depression in the zero-bias conductance is still observed at 
$T \simeq T_c$.  This, together with the slight asymmetry 
in the  $G$-$vs$-$V$ curves, can be linked to a residual barrier  
at the buried InGaAs/GaAs heterojunction.

In Fig. 3 we show how the conductance can be strongly modified by 
very weak magnetic fields ($H$) applied perpendicularly to the plane of the 
junction.  The $G$-$vs$-$V$ curves shown in Fig. 3 were
taken at $T=0.33$ K for different values of $H$ in the 
0--5 mT range. The superconducting gap vanishes for  
$H$ approaching the critical field ($H_c$) of 
the Al film ($H_c \simeq 10$ mT at $T=0.33$ K). 
Consequently, the zero-bias conductance dip is less and less
pronounced and at the same time shrinks with increasing magnetic field.
The latter effect was not as noticeable in Fig. 2 owing to the 
temperature-induced broadening of single-particle Fermi distribution 
function \cite{btk}. 

In conclusion, we have reported on Andreev-reflection dominated transport
in MBE-grown Si-engineered 
Al/n-In$_{0.38}$Ga$_{0.62}$As hybrid junctions. 
Transport properties
were studied as a function of temperature and magnetic field and 
showed junction transmissivity close to the theoretical limit for 
the S-Sm combination.
The present study demonstrates that the 
Si-interface-layer technique is a promising tool to 
obtain high-transparency S-Sm junctions involving InGaAs
alloys with low In content and low doping concentration.    
This technique yields Schottky-barrier-free junctions
without using InAs-based heterostructures and can be 
exploited in the most widespread MBE systems. It is 
particularly suitable for the realization of 
low-dimensional S-InGaAs hybrid systems grown on GaAs or InP substrates.
We should finally like to stress that its application in principle is not 
limited to Al metallizations and other superconductors 
could be equivalently used. In fact, to date the most convincing
interpretation of the silicon-assisted Schottky-barrier engineering is
based upon the heterovalency-induced IV/III-V local interface dipole
\cite{bin}. Within this description Schottky-barrier tuning is
a metal-independent effect.

The present work was supported by INFM under the PAIS project 
Eterostrutture Ibride Semiconduttore-Superconduttore. 
One of us (F. G.) would like to acknowledge Europa Metalli 
S.p.A. for financial support. 


\begin{figure}
\caption{ 
Schematic structure of the Al/n-In$_{0.38}$Ga$_{0.62}$As junctions studied in
this work. Further details are given in the text.
} 
\end{figure}

\begin{figure}
\caption{
Differential conductance $vs$ bias voltage 
of a Si-engineered Al/n-In$_{0.38}$Ga$_{0.62}$As single junction.
The four curves shown were obtained at zero magnetic field 
and temperatures in the  0.33--1.03 K range.
}
\end{figure}

\begin{figure}
\caption{
Differential conductance $vs$ bias voltage 
of a Si-engineered Al/n-In$_{0.38}$Ga$_{0.62}$As single junction.
The four curves shown were obtained at $T=0.33$ K 
under different magnetic fields perpendicular to the 
junction plane. 
} 
\end{figure}


\begin{references}

\bibitem{been} C. W. J. Beenakker, Rev. Mod. Phys. {\bf 69}, 731 (1997). 

\bibitem{klei} A. W. Kleinsasser and W. L. Gallagher, {\it
Superconducting Devices}, edited by S. Ruggiero and D. Rudman (Academic,
Boston, 1990), p. 325.

\bibitem{lamb} C. J. Lambert and R. Raimondi, J. Phys. Condens. Matter
{\bf 10}, 901 (1998). 

\bibitem{andreev} A. F. Andreev, Zh. Eksp. Teor. Fiz. {\bf 46},
1823 (1964) [Sov.Phys.--JETP {\bf 19}, 1228 (1964)].

\bibitem{kast} A. Kastalsky, A. W. Kleinsasser, L. H. Greene, R. Bhat,
F. P. Milliken, and J. P. Harbison, Phys. Rev. Lett. {\bf 67}, 3026
(1991).

\bibitem{nguy} C. Nguyen, H. Kroemer, and E. L. Hu, Appl. Phys. Lett.
{\bf 65}, 103 (1994).

\bibitem{akaz} T. Akazaki, J. Nitta, and H. Takayanagi, Appl. Phys.
Lett. {\bf 59}, 2037 (1991).

\bibitem{gao} J. R. Gao, J. P. Heida, B. J. van Wees, S. Bakker, and T.
M. Klapwijk, Appl. Phys. Lett. {\bf 63}, 334 (1993).

\bibitem{williams} A. M. Marsh, and D. A. Williams, J. Vac. Sci.
Technol. A {\bf 14}, 2577 (1996).

\bibitem{tabo} R. Taboryski, T. Clausen, J. Bindslev Hansen, J. L.
Skov, J. Kutchinsky, C.B. S{\o}rensen, and P. E. Lindelof, Appl. Phys.
Lett. {\bf 69}, 656 (1996).

\bibitem{silv} S. De Franceschi, F. Beltram, C. Marinelli,
L. Sorba, M. Lazzarino, B. M\"uller, and A. Franciosi, Appl. Phys. Lett.
{\bf 72}, 1996 (1998).

\bibitem{nota} 
The voltage drop across the junction amounts to about half of the
applied bias, due to the series-resistance contribution. 

\bibitem{van} W. M. van Huffelen, T. M. Klapwijk, D. R. Heslinga, M. J.
de Boer, and N. van der Post,  Phys. Rev.
B {\bf 47}, 5170 (1993).

\bibitem{klei} A. W. Kleinsasser, T. N. Jackson, D. McInturff, F. Rammo,
G. D. Pettit, and J. M. Woodall, Appl. Phys. Lett.
{\bf 57}, 1811 (1990).


\bibitem{chau} S. Chaudhuri and P. F. Bagwell, Phys. Rev.
B {\bf 51}, 16936 (1995).

\bibitem{btk} G. E. Blonder, M. Tinkham, and T. M. Klapwijk, Phys. Rev.
B {\bf 25}, 4515 (1982).

\bibitem{BT} G. E. Blonder and M. Tinkham, Phys. Rev. B {\bf 27}, 112
(1983).

\bibitem{bin} C. Berthod, J. Bardi, N. Binggeli, A. Baldereschi,
 J. Vac. Sci.  Technol. B {\bf 14}, 3000 (1996); C. Berthod, 
{\it et al.}, Phys. Rev. B {\bf 57}, 9757 (1998).  



\end{references}
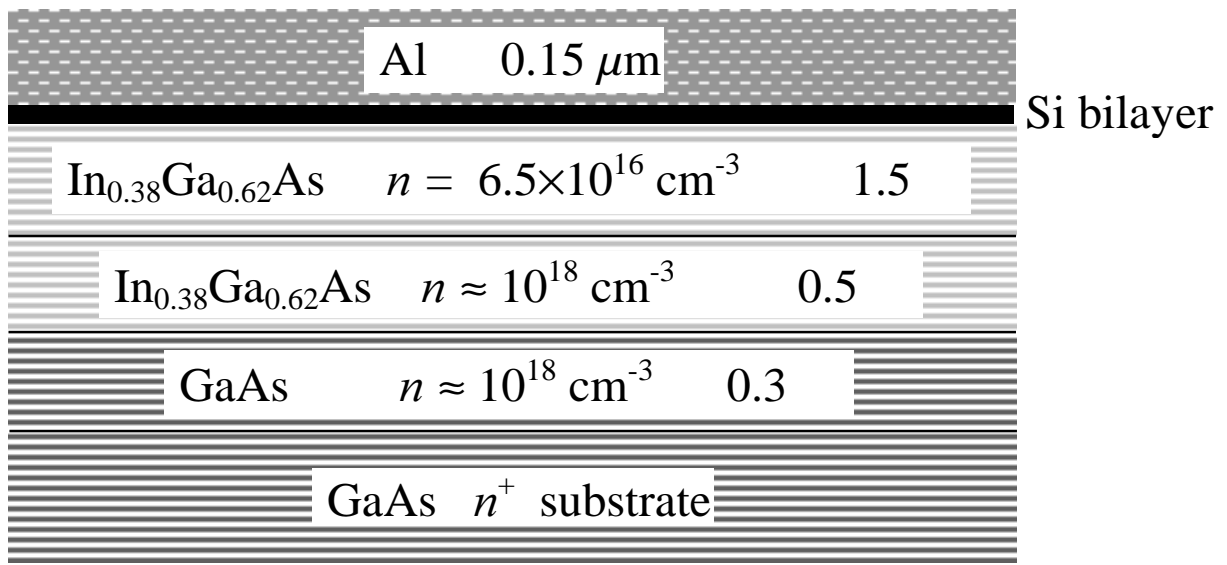
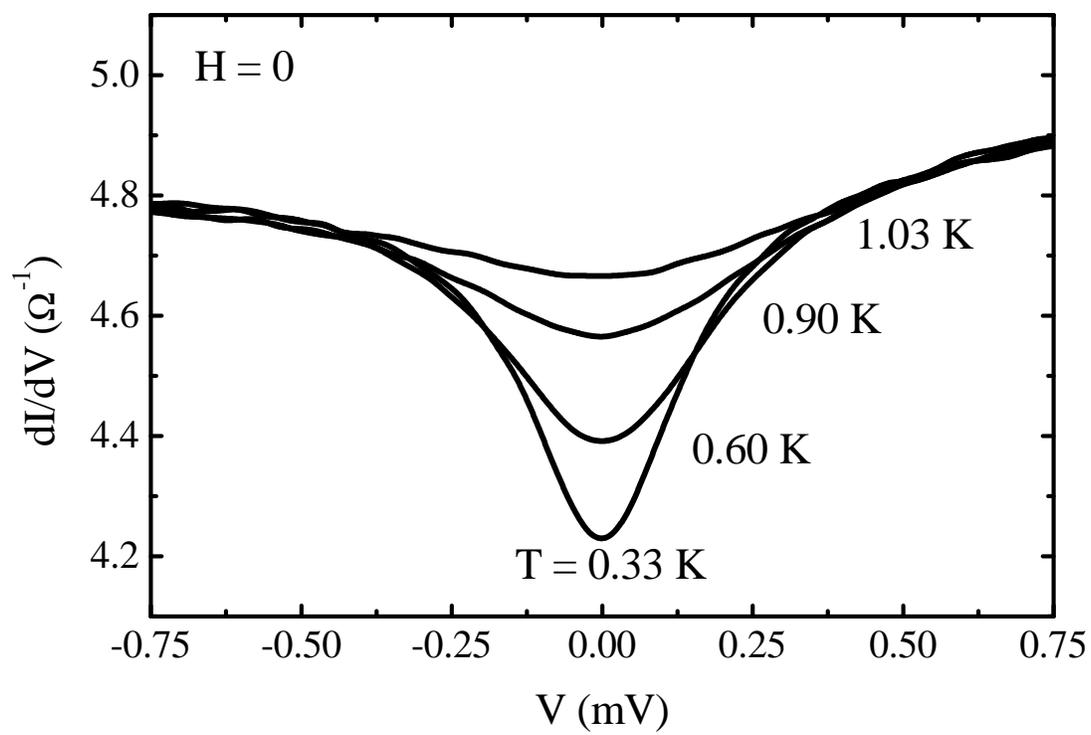
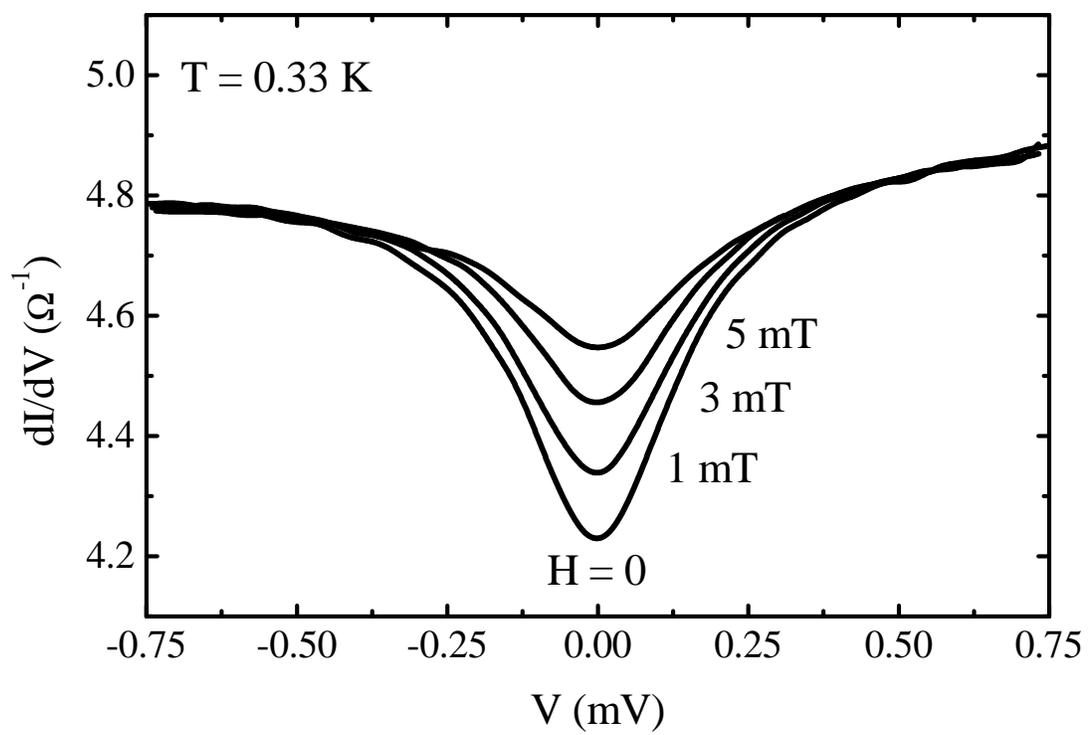
\end{document}